\documentclass[9pt]{article}
\begin{document}  
\title{NON-SINGULAR INHOMOGENEOUS STIFF 
FLUID COSMOLOGY}  
\author{L. Fern\'andez-Jambrina \footnote{Permanent address: 
Departamento de Geometr\'{\i}a y Topolog\'{\i}a, 
Facultad de Ciencias Matem\'aticas, 
Universidad Complutense de Madrid, 
E-28040-Madrid, Spain}  \\ Theoretisch-Physikalisches Institut\\Max-Wien-Platz 
1\\Friedrich-Schiller-Universit\"at-Jena\\07743-Jena, Germany}
\date{\empty}
 
\maketitle
\begin{abstract} 
In this talk we show a stiff fluid solution of the Einstein equations for a cylindrically 
symmetric spacetime. The main features of this metric are that it is non-separable in 
comoving coordinates for the congruence of the worldlineS of the fluid and that it yields 
regular curvature invariants. \end{abstract}

\section{Introduction}
The publication in 1990 of the first known cosmological perfect fluid
solution of the Einstein equations with regular curvature invariants
[1] has triggered the search of new solutions sharing this feature and
the analysis of their properties.  In a subsequent paper [2] it was
proven that this solution not only has no curvature singularity but it
is geodesically complete and singularity-free [3].  It has also been
shown that it is part of a larger family which includes both singular
and non-singular metrics [4].  Since then other non-singular solutions
have been obtained, all of them describing spacetimes with an abelian
$G_2$ group of isometries.  The main aim of this field of research would
be to establish whether there is an open set of metrics which are
non-singular.

In this talk we shall show a new diagonal metric which has got regular
curvature invariants and is causally stable.  It also possess an
abelian orthogonally transitive $G_2$ group of isometries.  The matter
source in this spacetime is a stiff perfect fluid.  To our knowledge
it is the first solution with cylindrical symmetry which is both
non-separable in comoving coordinates and non-singular .

\section{The line element}
For the line element constructed from the metric, $g$, of the solution
we choose a set of coordinates $\{t, \phi, r, z\}$, 
\begin{equation}
-\infty < t, z <\infty,\quad 0 < r < \infty,\quad 0 < \phi < 2\pi,
\end{equation} 
where the coordinates, $z$, and $\phi$, are adapted to the commuting
Killing vectors, so that the metric functions depend only on the
remaining coordinates, $r$, $t$, which are an isothermal
parametrization for the two-dimensional submanifolds $z =
\textrm{const}.$, $\phi =\mathrm{const.}$.  In this chart the metric
takes the form,
 \begin{equation}\label{metric}
ds^2={\rm e}^{ K(t,r)}\,(-dt^2+dr^2)+{\rm e}^{-U(t,r)}\,dz^2+
 {\rm e}^{U(t,r)}\,r^2\,d\phi^2,
\end{equation}
where the functions $U(t,r)$ and $K(t,r)$ that have been introduced
have the following expressions,
\begin{equation}
K(t,r)=\frac{1}{2}\,{ \beta}^{2}\,{r}^{4 } + ( \alpha+\beta)\,{r}^{2} +
2\,{t}^{2}\,{ \beta}
 + 4\,{t}^{2}\,{ \beta}^{2}\,{r}^{2}
 \end{equation}
 \begin{equation}
 U(t,r)={ \beta}\,(\,{r}^{2} + 2\,{t}^{2}\,).
 \end{equation}
The metric satisfies the regularity conditions in the vicinity of the
sub manifold $r = 0$ [5] and therefore it can be considered as an actual
symmetry axis.  A restriction needs be imposed on the values of the
only free parameters $\alpha$ and $\beta$, 
\begin{equation}\alpha>0,\qquad\beta>0.\end{equation}

\section{Curvature invariants}
In order to write the expressions for the components of the Weyl
tensor we shall make use of a complex null tetrad,
\begin{equation}
l=\frac{\theta^0+\theta^1}{\sqrt{2}},\quad
n=\frac{\theta^0-\theta^1}{\sqrt{2}},\quad
m=\frac{\theta^2+i\theta^3}{\sqrt{2}},\quad
\bar m=\frac{\theta^2-i\theta^3}{\sqrt{2}},
\end{equation}
that can be constructed from the orthonormal coframe 
$\{\theta^0,\theta^1,\theta^2,\theta^3\}$,, 
\begin{equation}
\theta^0={\rm e}^{\frac{1}{2}\,K(t,r)}\,dt,\ \ \ \theta^1={\rm
e}^{\frac{1}{2}\,K(t,r)}\,dr, \ \ \ \theta^2={\rm e}^{-\frac{1}{2}
U(t,r)}\,dz, \ \ \ \theta^3=
 {\rm e}^{\frac{1}{2}U(t,r)}\,r\,d\phi.\label{coframe}
\end{equation}

After introducing two functions, $f_{1}$, $f_{2}$, 
\begin{equation}
f_1(t,r)=- 3\,{ \beta}
 + 2\,{ \beta}^{3}\,{r}^{4} + 3\,{ \beta}^{2}\,{r}^{2} +  \alpha
\,(1+2\,\beta\,r^2) + 24\,{ \beta}^{3}\,{r}^{2}\,{t}^{2} +12\,{ 
\beta}^{2}\,{t}^{2}
\end{equation}
\begin{equation}
f_2(t,r)=12\,{ \beta}^{3}\,{r}^{3}\,{t} + 12\,{ 
\beta}^{2}\,{r}\,{t} + 16\,{ \beta}^{3}\,{t}
^{3}\,{r} + 4\,{ \alpha} \,{ \beta}\,{t}\,{r}.
\end{equation}
the components of the Weyl tensor in the null tetrad can be written as follows, 
\begin{equation}
\Psi_0=\frac{1}{2}\left(f_1(t,r)+f_2(t,r)\right) \,{\rm e}^{ 
-K(t,r) }
\end{equation}

\begin{equation}
\Psi_1=0
\end{equation}
\begin{eqnarray}
{ \Psi_2}= {\displaystyle \frac {1}{6}}\, \left( \! \,  3\,{
 \beta}^{2}\,{r}^{2} + 3\,{ \beta} - { \alpha} - 12\,{ 
\beta}^{2}\,{t}^{2}\, \!  \right) \,{\rm e}^{ -K(t,r) }
\end{eqnarray}
\begin{equation}
{ \Psi_3}=0
\end{equation}
\begin{equation}
\Psi_4=\frac{1}{2}\left(f_1(t,r)-f_2(t,r)\right) \,{\rm e}^{ 
-K(t,r) }.
\end{equation}

\section{The energy momentum tensor}
The energy momentum tensor corresponds to a stiff perfect fluid, 
\begin{equation}
T=\mu u\otimes u +p(g+u\otimes u),
\end{equation}
where the pressure, $p$, is equal to the density, $\mu$, of the fluid 
\begin{equation}
\mu=p= \alpha\,{\rm e}^{-K(t,r)},
\end{equation}
and happens to be non-singular.  This metric can be generated from a
vacuum spacetime making use of the Wainwright, Ince, Marshman
algorithm [6].  The seed metric is obtained taking the zero value for
$\alpha$ in (2).  

The four-velocity of the fluid has only projection on $\partial_{t}$,
\begin{equation} {u}= {\rm e}^{ -\frac{1}{2}\,K(t,r)}\partial_t,
 \end{equation}
since the coordinates that have been chosen are comoving. 
This implies that the acceleration of the fluid has only a radial component, 
\begin{equation}
a= {r}\, \left( \! \,{ \beta}^{2}\,{r
}^{2} + { \alpha} + { \beta} + 4\,{ \beta}^{2}\,{t}^{2}\, \! 
 \right) \, \partial_r,
\end{equation}
due to the orthogonal transitivity requirement and to the fact that
the velocity is orthogonal to the orbits of the group of isometries.
The only fluid wordlines that are geodesic are those contained in the
$r = 0$ submanifold.

The shear tensor for the fluid has the following expression, 
\begin{equation}
\sigma=\frac{4}{3}\,{ \beta
}\,{t}\,{\rm 
e}^{-\frac{1}{2}\,K(t,r)}\,\left\{(\,1 + 2\,{ 
\beta}\,{r}^{2}\,)\,\theta^1\otimes \theta^1-(\,2 + { 
\beta}\,{r}^{2}\,)\,\theta^2\otimes \theta^2+(1-\,{ \beta}\,{r}^{2
})\,\theta^3\otimes 
\theta^3\right\}.
\end{equation}
in the orthonorrnal coframe defined in (7).  There is no vorticity
since we are dealing with an orthogonally transitive $G_2$ group of
isometries.  Finally we write down the expansion, $\Theta$, of the
cosmological fluid,
\begin{equation}
{ \Theta}=2\,\beta\,t\,(\,1 + 2\,{ \beta}\,{r}^{2}\,)\,{\rm 
e}^{-\frac{1}{2}\,K(t,r)}, 
\end{equation}
which shows that the spacetime is contracting for negative values of
the time coordinate and expanding for positive values of it.

\section{Discussion}

From the expressions for the Weyl tensor and the fluid density it is
straightforward to check that there is no curvature singularity in
this spacetime, since the curvature invariants are polynomial
functions of them.

Besides the time coordinate, $t$, is a globally defined function whose
gradient is always timelike, that is, it is a cosmic time [3] and
therefore the spacetime is causally stable and the hierarchy of
causality conditions under it (strong causality condition, chronology
condition, ...  ) are satisfied.

The strong and dominant energy conditions [3] hold since the energy
density of the fluid is positive everywhere and the equation of state
corresponds to a stiff fluid.  Also the generic condition on the
Riemann tensor is fulfilled, since the fluid density does not vanish
[7].

It remains to be shown whether the spacetime is bundle or geodesicaJly
complete.  If the latter were true, the reason for avoiding the
existence of singularities, according to the singularity theorems [3],
would have to be looked for in the lack of closed trapped surfaces or
of compact achronal sets without edge or of a null geodesic focalizing
point in the spacetime, as it happens in other regular spacetimes [2].
There is work in progress in this direction.  It would also be
interesting to prove global hyperbolicity.

The curvature tends to zero for large values of the radial, $r$, and
time, $t$, coordinates, as it can be inferred from the expressions for
the components of the Weyl tensor and the density and the pressure of
the fluid.  The fluid is rather diluted in early times and becomes
more and more dense until the time coordinate reaches the zero value.
At this time the universe starts to expand.  This feature is typical
in other singularity-free cosmological models [1].  As in every other
known non-singular cosmological model, the gradient of the
transitivity surface area element of the metric is spacelike.  There
is a static limit for vanishing $\beta$ which happens to be same as for the
non-singular metric in [8].

Further details on this spacetime will be given in a forthcoming 
paper [9].

\vspace{0.3cm} \noindent{\textbf{Acknowledgements.} The present work
has been supported by the Direccion General de Enseiianza Superior
Project PB95-0371 and by a DAAD (Deutscher Akademischer Austausch-
dienst) grant for foreign scientists.  The author wishes to thank
Prof.  F. J. Chinea and Dr.  L. M. Gonzalez-Romero for valuable
discussions and Prof.  Dietrich Kramer and the Theoretisch-
Physikalisches Institut of the Friedrich-Schiller-Universitiit-Jena
for their hospitality.

\end{document}